\documentclass[journal,twoside,web]{ieeecolor}
\usepackage{generic}
\usepackage{cite}
\usepackage{amsmath,amssymb,amsfonts}
\usepackage{algorithmic}
\usepackage{graphicx}
\usepackage{algorithm,algorithmic}
\usepackage{hyperref}
\hypersetup{hidelinks=true}
\usepackage{textcomp}

\usepackage{amsmath}

\usepackage{multirow}
\usepackage[table,xcdraw]{xcolor}

\usepackage{etoolbox}
\makeatletter
\patchcmd{\@makecaption}
  {\scshape}
  {}  
  {}
  {}
\makeatother

\usepackage{amsmath} 

\usepackage{graphicx}

\usepackage{cite}
\usepackage{amsmath,amssymb,amsfonts}
\usepackage{algorithmic}
\usepackage{graphicx}
\usepackage{textcomp}
\usepackage{xcolor}
\def\BibTeX{{\rm B\kern-.05em{\sc i\kern-.025em b}\kern-.08em
    T\kern-.1667em\lower.7ex\hbox{E}\kern-.125emX}}

\def\BibTeX{{\rm B\kern-.05em{\sc i\kern-.025em b}\kern-.08em
    T\kern-.1667em\lower.7ex\hbox{E}\kern-.125emX}}
\begin{document}
\title{HAGAN: Hybrid Augmented Generative \\ Adversarial Network for Medical \\ Image Synthesis}

\author{Zhihan Ju, Wanting Zhou$^{{*}}$, Longteng Kong, Yu Chen, Yi Li, Zhenan Sun, Caifeng Shan
\thanks{This work was supported in part by the National Natural Science Foundation of China (Grant No.62376037, 62006227, 82202244), and the Open Project Program of the Key Laboratory of Artificial Intelligence for Perception and Understanding, Liaoning Province (AIPU)(No.20230006).
 ($^{{*}}$Wanting Zhou is the corresponding author)}
\thanks{Z. Ju, W. Zhou and L. Kong are with Beijing University of Posts and Telecommunications, Beijing 100876, China (e-mail: juzhihan@bupt.edu.cn; wanting.zhou@bupt.edu.cn; konglongteng@bupt.edu.cn).}
\thanks{Y. Chen is with Beijing Tiantan Hospital,Capital Medical University, Beijing 100070, China (e-mail: chenyu\_tiantan@126.com).}
\thanks{Y. Li is with Dalian University of Technology, Liaoning 116081, China(e-mail: liyi@dlut.edu.cn).}
\thanks{Z. Sun is with Institute of Automation, Chinese Academy of Sciences, Beijing 100190, China (znsun@nlpr.ia.ac.cn).}
\thanks{C. Shan is with the School of Intelligence Science and Technology, Nanjing University, Nanjing 210023, China. (Email: caifeng.shan@gmail.com)}}

\maketitle

\begin{abstract}
Medical Image Synthesis (MIS) plays an important role in the intelligent medical field, which greatly saves the economic and time costs of medical diagnosis. However, due to the complexity of medical images and similar characteristics of different tissue cells, existing methods face great challenges in meeting their biological consistency. To this end, we propose the Hybrid Augmented Generative Adversarial Network (HAGAN) to maintain the authenticity of structural texture and tissue cells. HAGAN contains Attention Mixed (AttnMix) Generator, Hierarchical Discriminator and Reverse Skip Connection between Discriminator and Generator. The AttnMix consistency differentiable regularization encourages the perception in structural and textural variations between real and fake images, which improves the pathological integrity of synthetic images and the accuracy of features in local areas. The Hierarchical Discriminator introduces pixel-by-pixel discriminant feedback to generator for enhancing the saliency and discriminance of global and local details simultaneously. The Reverse Skip Connection further improves the accuracy for fine details by fusing real and synthetic distribution features. Our experimental evaluations on three datasets of different scales, i.e., COVID-CT, ACDC and BraTS2018, demonstrate that HAGAN outperforms the existing methods and achieves state-of-the-art performance in both high-resolution and low-resolution.  

\end{abstract}


\section{Introduction}
With the development of medicine, more and more diseases can be detected at an early stage,\begin{figure}[htp]
    \centering
    \includegraphics[width=8.8cm]{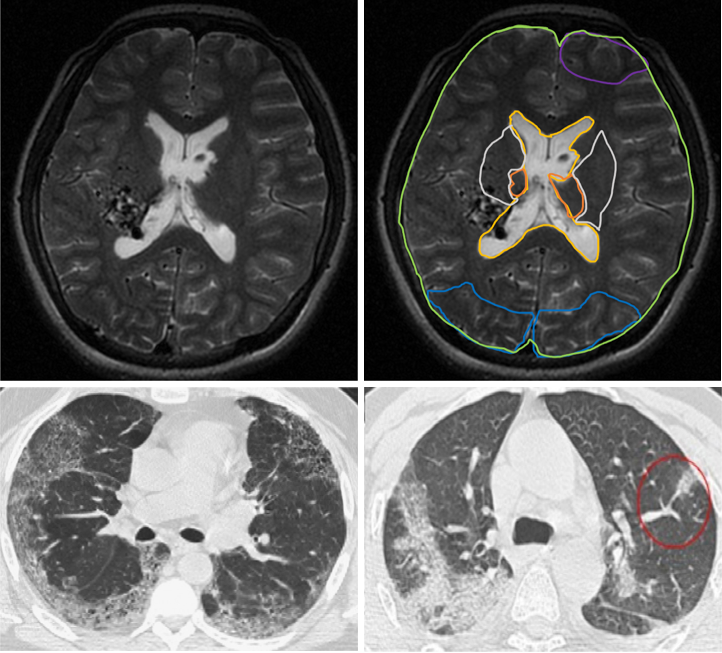}
    \caption{Illustration of the real medical imaging samples in lungs and brain. The structural texture features of medical images are very complex and subtle. Moreover, the characteristics of tissue cells in medical imaging are similar. It is challenging to distinguish the feature disparities between key lesions and different tissues.}
    \label{fig:0}
\end{figure} relying on top experts to capture subtle lesions or abnormal representations of tissues in medical imaging. However, such judgments rely heavily on doctors' expertise, and the data collection process also requires a lot of manpower, time, and funding. In order to improve the quality of medical imaging and reduce the cost of high-precision acquisition, computer experts have tried to use generative adversarial networks (GANs) to synthesize high-quality medical images. This includes unconditional image synthesis \cite{21,22,23} and transitions between different states of the image, such as cross-domain conversion of medical image in different modes\cite {16,74,75,76}, high-quality reconstruction of low-dose Computed Tomography (CT)\cite {8,77,78,79,80} or low-quality Magnetic Resonance Imaging (MRI)\cite {11,81,82,83,84}, and 3D synthesis of medical slice\cite {4,89,90}. Among them, more and more experts are committed to researching unconditional synthesis technology, which can fundamentally improve the scale and diversity of medical image datasets. Unconditional medical image synthesis technique does not require any conditional information. The generative model will learn the data distribution characteristics of real images and synthesize medical images from its latent variable space. How to more effectively ensure the authenticity of medical synthetic images is the main proposition of this technology. The synthesized medical images are often used for pathology training of specialized physicians or as training data to enhance the performance of models on medical downstream tasks, such as medical image segmentation and pathological classification models. 

\par In recent years, the quality of medical images synthesized by GANs has been greatly improved, and the reasoning ability of models have been gradually enhanced. Experts improve the generalization ability by improving the architecture of GANs\cite{6, 21, 65, 71, 86, 69, 70}. And some researchers use denoising diffusion models for medical image synthesis\cite{70, 87, 88}, while others are trying to use the transformer architecture to improve the ability of extracting long-distance contextual information, thereby improving the generation of overall pathological structures\cite{35, 66}. In addition, they try to generate high-resolution medical images to obtain finer local details\cite{3, 85}. And the stability of model training has been effectively improved through various regularization techniques \cite{46, 55, 56}. Although experts have designed diverse structures and training methods for medical image synthesis, the data representation capabilities of existing models still face great challenges in capturing the features of medical images.

\par Firstly, medical images often have complex and extremely subtle structural texture features. For existing medical image synthesis methods, they are difficult to constrain the model to pay extra attention to the integrity of the synthesized image structure texture beyond key distinguishing features by only discriminating between the real and fake images. The minor generation deviation will seriously affect the histopathological characteristics of the image and undermine its biological authenticity. Secondly, even on the same medical image, the key features or lesion information between different tissues are very small and difficult to capture, such as the brain MRI in Figure 1, where the characteristics of different tissue cells are very similar. Even for important visual and language centers, it is difficult to find key distinguishing features in medical imaging. In addition, in the lung CT image of Figure 1, the COVID-19 lesion is only presented in the red box, and its features are more difficult to capture. For existing methods, they only focus on the true and fake features at global image level, and lack the feature feedback of local details, making the models difficult to pay attention to the imaging differences of local tissue cells, and thus unable to synthesize real and accurate local tissues or lesion sites.

\par Therefore, in order to maintain the pathological structural integrity and local detail consistency of the synthesized images, we propose the Hybrid Augmented Generative Adversarial Network (HAGAN) for medical image synthesis. HAGAN contains three main modules , namely Attention Mixed (AttnMix) Generator, Hierarchical Discriminator and Reverse Skip Connection between Discriminator($D$) and Generator($G$). For improving the pathological integrity of the synthesized images,  AttnMix consistency differentiable regularization constrains the model to be perceptive in the structural and textural variations between real and fake images. The success of UNetGAN\cite{30} also proves the effectiveness of consistency regularization methods in enhancing model data representation capabilities. Furthermore, AttnMix introduces the visual self-attention mechanism to promote the model more focused on key features of tissue cells or lesion sites at zero cost through attention maps. To ensure the biological consistency of global and local details, Hierarchical Discriminator provides discriminant feedback at both the image and pixel bi-level to the generator for improving the saliency and discriminance of the model. Moreover, we propose Reverse Skip Connection structure, which improves local localization accuracy and accelerates generator convergence by fusing feature maps under the true discriminant path into the generated path. Finally, we improve the quality of structural texture and local tissue in synthetic medical images through joint learning based on the above methods.

\par Our contributions could be summarized as follows.
\begin{itemize}
\item We propose the Hybrid Augmented Generative Adversarial Network (HAGAN), which improves structural integrity of pathology and consistency of local details through bi-level consistency constraints and architecture modification. 

\item We propose the Attention Mixed(AttnMix) Generator, which resorts to the consistency differentiable regularization to ensure model to be perceptive in the structural and textural information of medical images. Moreover, we propose the Hierarchical Discriminator with Reverse Skip Connection to enhance the saliency and discriminability of feature extraction. 

\item We propose the image and pixel bi-level adversarial loss and consistency regularization loss, which can lead to profound improvement of model performance.

\item HAGAN is applied to medical image synthesis and achieved state-of-the-art performance at low and high resolution on COVID-CT, ACDC and BraTS2018 datasets of three different scales.
\end{itemize}

\section{Related Work}
\textbf{Medical Image Synthesis.} Generative Adversarial Networks (GANs) have been deeply applied in the past few years due to excellent performance. In the field of medical image synthesis, GANs have demonstrated significant effectiveness. They have been utilized for a variety of tasks, including data amplification \cite{1, 2, 3, 6, 7, 21, 23, 35}, multi-contrast MRI synthesis \cite{10, 11, 12, 13, 14, 15}, low-dose CT image reconstruction \cite{8, 9},  3D MRI reconstruction \cite{4, 5} and image cross-modal synthesis \cite{16, 17, 18, 19, 20}. For instance, \cite{68} used deep convolutional generative adversarial network(DCGAN) to create artificial images for improving the performance of chest pathology classification. \cite{69} implemented a Wasserstein GAN (WGAN) for generating synthetic multi-sequence brain Magnetic Resonance images. \cite{21} trained a GAN to synthesize new T1 brain MRIs. \cite{22} successfully generated High-resolution skin lesion images that experts could not distinguish from real images. Additionally, GANs were used in another effort \cite{23} to generate both qualitative and high-resolution MRIs of the retinal fundus. And \cite{3} used the PGGAN model to iteratively generate high-resolution composite images of MRI and retinal fundus with sizes of 1024 × 1024. In \cite{1, 2}, authors demonstrated that images of lung cancer nodules generated by GANs are almost indistinguishable from real images. These synthetic images proved difficult to differentiate even for trained radiologists. Medical image synthesis techniques are widely used, but they are limited by the scale of medical datasets, the instability of training and the requirement for pathological structural integrity and local tissue consistency. As a result, the synthesized images produced by such methods often suffer from incomplete pathological structures and discontinuities in local details, as shown in Figures 5 and 6. People have made a lot of attempts to improve the performance of GANs around the network structure. For example, SAGAN \cite{28} added a visual self-attention mechanism to generator for improving the understanding of global structural information; UNetGAN\cite{30} uses UNet structure as the discriminator and achieves pixel-wise discriminator feedback by progressively restoring the downsampling results. In addition, people also try to add the Vision Transformer\cite{31} structure to the generative adversarial network, and use its processing ability for long-distance information to improve the capture of global information. For example, ResVit \cite{35} uses the sensitivity of transformer blocks to process context information and the accuracy of convolution operators to propose a new type of medical image synthesis network. At the same time, diffusion probabilistic models have been increasingly used to generate high-quality images, \cite{70} used denoising diffusion probabilistic models (DDPM) to synthesize high quality histopathology images of brain cancer. In this paper, we propose an independently decoupled image and pixel bi-level discriminator, while providing structural integrity and detail continuity constraints for discrimination. Moreover, we construct Reverse Skip Connection between generated and discriminated paths to better capture global and local information.

\par \textbf{Data Augmentation.} GANs are very dependent on the diversity and breadth of data. In the field of medical images, collecting high-quality large-scale datasets requires a lot of manpower and material resources, and involves data sensitivity, privacy and other security issues. In the field of image recognition, traditional data augmentation methods achieve amplification of training data through class domain preservation transformations, such as color transformation \cite{37}, area masking \cite{38}, flipping, rotation, cropping \cite{39}, data mixing \cite{40} and local and bionic distortion \cite{41}. However, these traditional augmentation methods are not suitable for generative adversarial networks. Augmenting real samples alone or inconsistent random augmentation will make the generator more inclined to match the data distribution of the augmented images, which will cause serious distribution drift. At the same time, real samples and fake samples for data augmented operations will also lead to completely different convergence directions for the generator and discriminator. \cite{36} proposed a method to perform the same type of differentiable data augmented operation on the real and fake samples at the same time, so that the gradient to be backpropagated from the augmented part of the image to generator and achieve regularization on the discriminator. Through qualitative and quantitative experiments on small-scale datasets, it is proved that the method can greatly alleviate the overfitting problem of generative adversarial networks without reducing generative diversity. In this work, we also introduce similar differentiable data augmentation methods to improve the generation performance of the model.

\begin{figure*}[htp]
    \centering
    \includegraphics[width=18.2cm]{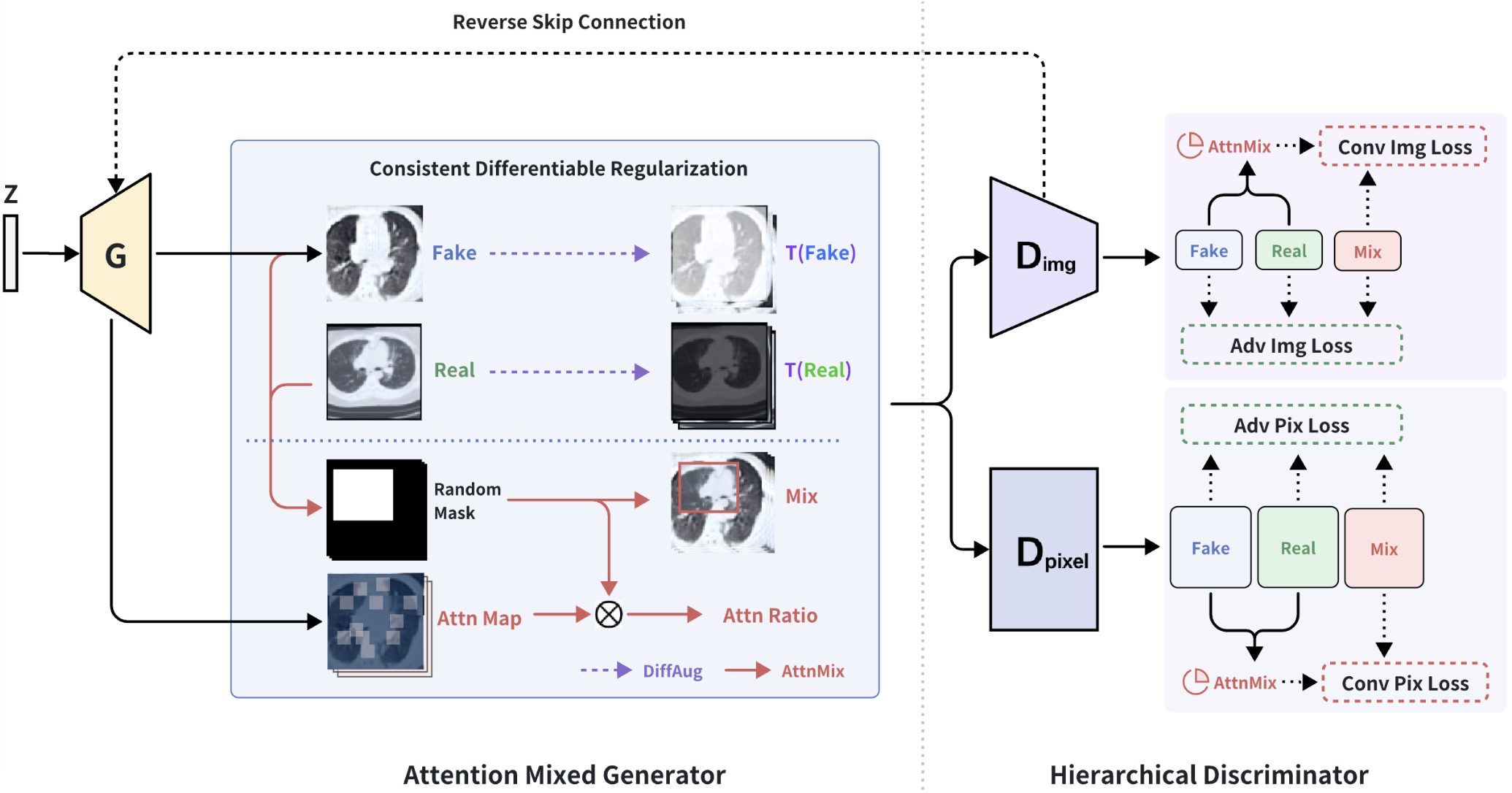}
    \caption{Overview of proposed Hybrid Augmented Generative
Adversarial Network(HAGAN). Given a random noise z, first synthesize the medical image through the generator, and use the AttnMix to complete the differentiable augmentation and mixing of real and fake images to ensure model to be perceptive in the structural and textural variations. After that, the T(fake), T(real) and Mix input to Hierarchical Discriminator, and calculates the bi-level adversarial loss and consistency loss to achieve joint learning to maintain structural integrity of pathology and consistency of local texture details.}
    \label{fig:2}
\end{figure*}

\par \textbf{Consistency Regularization.} In recent years, a large number of methods have tried to optimize model performance from a regularization perspective\cite{46}, which take advantage of the basic fact that if the same image is perturbed, the recognition results should be consistent. Simple and effective regularization techniques have been proposed to create composite images by mixing the samples of different categories. Among them, Mixup \cite{40} and CutMix \cite{43} are the two most common methods. Mixup obtains augmented images by pixel-weighted combination of two images in the numerical dimension; CutMix achieves spatial-level image augmentation by cropping and blending the same-position regions of the two images. However, neither of the above two methods can fully align the synthetic sample space with the label space, so saliency-based methods have been proposed, including puzzle-Mix \cite{45}, saliency-CutMix \cite{47} and TransMix \cite{48}. Transmix uses the feature map generated by self-attention mechanism of transformer structure to readjust the label space for aligning the synthetic distribution of the sample space. Due to the effectiveness of consistency regularization, UNetGAN\cite{30} introduces the technique into natural image synthesis for the first time. It uses CutMix to mix true and fake images for regularizing discriminator to improve generation performance. In this paper, we propose a more powerful consistency regularization method (AttnMix) based on visual self-attention mechanism to improve the data representation ability of the structure and texture features. And we apply consistency regularization technology to medical image synthesis for the first time.

\section{Methods}

\begin{figure}[htp]
    \centering
    \includegraphics[width=8.8cm]{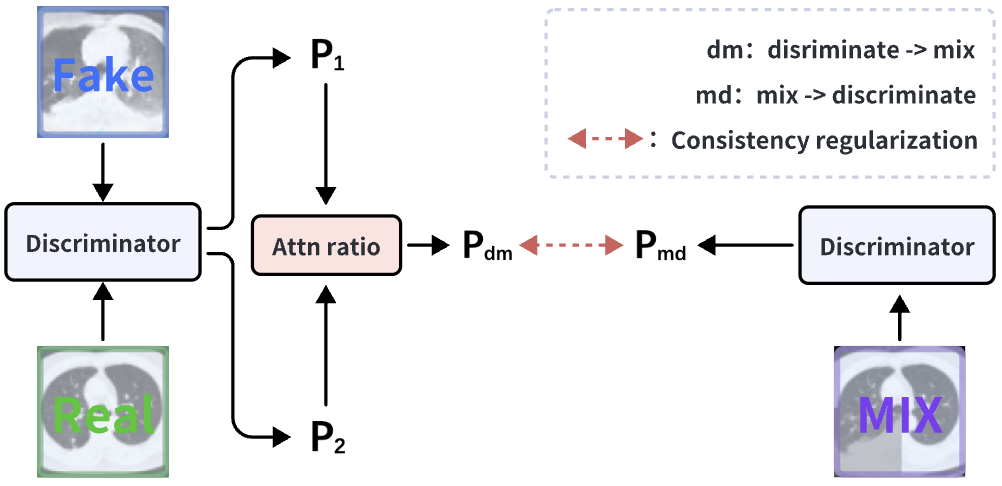}
    \caption{Illustration of AttnMix consistency differentiable regularization. P1 and P2 are the discrimination results of fake images and real images respectively. }
    \label{fig:1}
\end{figure}

\subsection{Overview}
The HAGAN method contains the Attention Mixed Generator, Hierarchical Discriminator with Reverse Skip Connection and the additional loss functions. The AttnMix Generator resorts to consistency differentiable regularization for ensuring model to be perceptive in structural and textural features. The Hierarchical Discriminator with Reverse Skip Connection further improves the global and local localization accuracy and enhances the saliency and discriminance of the model. The Loss functions introduce image and pixel bi-level adversarial loss and consistency loss into optimization goals to improve generalization performance. The complete model structure can refer to Figure 2. 

\subsection{Attention Mixed Generator}
The Attention Mixed Generator uses AttnMix consistent differentiable
regularization to solve two problems. First, well-trained generative models should pay more attention to structural and textural information, and less attention to arbitrary class domain preservation perturbations. Under the input of the same images, an equivalent result should be output. So we should further regularize the model to be perceptive in the structural and textural variations between real and fake images. This will enable generative models to pay more attention to meaningful structural and texture information, thereby improving the consistency of pathological structure and local details. Second, generative model training is extremely dependent on the diversity and breadth of data. Traditional data augmentation techniques enhance real samples alone or inconsistent random augmentation, resulting in serious distribution drift. Breaking through data bottlenecks in medical field requires matching data augmentation methods.

\par In this work, we propose AttnMix, a new hybrid augmented method. This method creates a composite image by cutting and pasting image patches at the same position under the true and fake categories, which can maintain structural integrity and distribution authenticity of medical images. At the same time, the self-attention graph generated in the training process of generator is used to generate the salient hybrid label, and the discriminator is constrained to be perceptive in the structural and textural variations between real and fake images without increasing any training cost. The Figure 3 and Figure 7 visually show the augmented strategy of Attnmix and the adjustment of label allocation.

\par First, we synthesize a new sample $i$ by mixing the generated sample $G (z)$ and the real sample $x$, where $M$ is a binary mask of the same size as $x$.

\begin{equation}
AttnMix(x, G(z), M) = M \odot x + (1-M) \odot G(z)
\end{equation}

\begin{equation}
i = AttnMix(x, G(z), M)
\end{equation}

\par Where the pixel value of $M$ is {0,1}, so if the pixel is derived from $G (z)$, the corresponding $M$ is 0, otherwise it is 1. Note that for the generated sample $i$, the label at the image level is the ratio of the cutting area to the original image area $\lambda$, and the label at the pixel level is $M$.

\par On this basis, Attnmix introduces the $A$ of the attention map generated by the last self-attention layer of the generator, and tries to adjust the ratio allocation of $\lambda$ based on the attention. This method aligns the sample space and the label space from the saliency dimension, solves the problem that mixed samples and labels are not aligned, and more accurately enhances the discrimination ability and constraints on spatial structure features learning.

\begin{equation}
\lambda_{1} = \sum_{i, j}(M\odot A) / \sum_{i, j}(M\odot A + (1-M)\odot A)
\end{equation}

\begin{equation}
\lambda = \alpha(\lambda_{0} + \lambda{1})
\end{equation}

\par Among them, $\lambda_{0}$ is the original area-based label allocation, $\alpha$ is the normalization parameter and $\lambda$ is the final image mixing ratio. In view of the AttnMix operation of the above formula, the generator generates real and fake images after differential augmentation, as well as mixed images. We need constrain the discriminator to provide consistent predicted values by adding the consistency loss to the discriminator target, which has image-level and pixel-level discrimination result consistency, as shown in Figure 3. The image level prompts the discriminator to learn the mixing ratio of true and fake images, further enhancing the discrimination ability of pathological integrity and local details, while the pixel level constrains the discriminator to learn structural and textural information. 

\begin{equation}
L_{cons} = || D(i) - mix(D(x) ,D(G(z),M))||^2
\end{equation}

\par In addition, we also use $i$ to train discriminator. In order to further improve the performance on small-scale medical image datasets, we introduce DiffAug technology, and perform differentiable augmentation($T$). So the data actually input to the discriminator participating in the training has three parts, namely $T (G (z))$, $T (x)$, and $i$. For HAGAN, we use an unsaturated target representation. In fact, the introduced HAGAN and the two-branch structure of the discriminator can be combined with any other adversarial loss.

\subsection{Hierarchical Discriminator}
\par In this paper, we propose a discriminator structure with two discriminant branches that are independently decoupled, where the discriminant branch at the image level is a traditional four-layer downsampling path that classifies the input image as true(1) or fake(0). The pixel branch is a three-layer channel dimension downsampling path that repeats the true-fake category judgment at the pixel level, dividing the image into true and fake regions. This structure enables the discriminator to learn global and local differences between true and fake images for enhancing the saliency and discriminance. 
\par Thereafter, we call the image-level discrimination branch $D^{img}$, and the introduced pixel-level discrimination branch $D^{pixel}$. Now, we calculate the loss using the image and pixel two-level discrimination.

\begin{equation}
L_{D} = L_{D}^{img} + \beta L_{D}^{pixel}
\end{equation}

\par Among them, $L_{D}^{img}$ and $L_{D}^{pixel}$ are the adversarial loss function commonly used in generative adversarial networks and $\beta$ is hyper-parameter that determines the ratio of $L_{D}^{pixel}$. $L_{D}^{img}$ mainly includes global structural integrity information, while $L_{D}^{pixel}$ includes local details of tissue and authenticity information.

\subsection{Reserve Skip Connection}
\par In the classic model U-Net in the field of image segmentation, it adopts Skip connections between Encoder-Decoder to improve the ability of segmenting fine details. Similarly, in this work, we treat generative adversarial networks as Decoder-Encoder architecture, whose inputs pass through the upsampling network first, and then pass through the downsampling network for discrimination.
\par Referring to the idea, we first add Reverse Skip Connection between the generator and discriminator of generative adversarial network. By postbacking the feature maps generated in the real image discrimination path to the corresponding resolution level of the generation path, feature fusion is realized, and the ability of the generator to accurately generate fine details is improved. At the same time, the training process is stabilized and convergence is accelerated, as shown in the Figure 4.

\begin{figure}[htp]
    \centering
    \includegraphics[width=8.3cm]{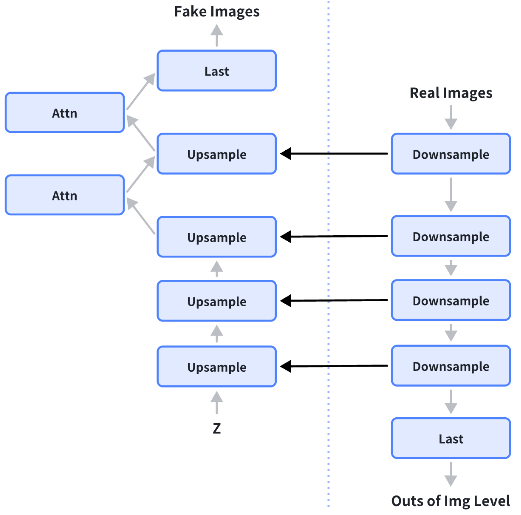}
    \caption{The structure of Reserve Skip Connection. The left is generator and the right is discriminator. }
    \label{fig:3}
\end{figure}

\subsection{Optimization and Training Strategies}
\par \textbf{Loss Functions.} In summary, the loss includes three parts: image-level adversarial loss, pixel-level adversarial loss, and consistency contrast loss. The optimization goal of the discriminator is as follows. $\beta_{1}$, $\beta_{2}$ are hyper-parameters to leverage the relative importance of different loss components. 

\begin{equation}
L_{D} = L_{D}^{img} + \beta_{1} L_{D}^{pixel} + \beta_{2} L_{cons}
\end{equation}

\par Correspondingly, the optimization goal of the generator has also become:
\begin{equation}
L_{G} = L_{G}^{img} + \beta L_{G}^{pixel}
\end{equation}

\par This encourages the generator to pay attention to both global structure and local details when synthesizing images, in order to fool more powerful discriminator.

\par We use the basic L2 loss function for the adversarial loss. At the same time, in the consistency regularization loss, we make relatively strict constraints on the consistency of the feature maps and results, so as to constrain the discriminator to ignore the category preservation disturbance, and further focus on structural and textural information. Among them, for the constraint of feature map consistency, we use contrast loss to compare the feature map extracted from the mixed image with the feature map of the corresponding level of real or fake images for consistency. In order to maintain uniformity, we use the same L2 loss function as the adversarial loss to ensure the consistency of the discrimination results.
\\
\par \textbf{Training Strategies.} This paper also adopts a simple and effective training strategy that integrates global prior information. We regard the training process as two parts, i.e., conventional training and differentially augmented training. The complete real data is used for preliminary learning to extract prior information with a true and complete pathological structure. Then, the augmented data through AttnMix is used to obtain stronger processing capabilities for local lesion details. In this way, the negative training impact that may be caused by data augmentation is avoided, and the training stability is improved.

\section{Experiments}

\subsection{Experimental Setup}
\par \textbf{Datasets.} In this paper, we conducted cross-sectional comparison experiments of the methods using three public medical image datasets, COVID-CT \cite{58}, ACDC \cite{59}, and BraTS2018 \cite{60}, which are of different sizes. In order to meet the different data collection and feature recognition needs of mobile devices and professional devices, the methods was tested on the resolution of 64 × 64 and 256 × 256. \textbf{COVID-CT}\cite{58} dataset contains 349 COVID-19 CT images from 216 patients and 463 non-COVID-19 CT images. The utility of the dataset is confirmed by a senior radiologist who has intensively practiced diagnosis and treatment of COVID-19 patients. In this experiment, we only used 396 non-COVID-19 CT images for training the models. \textbf{ACDC}\cite{59} dataset is composed of 2-dimensional heart cineMRIs from 100 patients. The cineMRIs were obtained using two MRI scanners of various magnetic strengths and different resolutions. In this experiment, we divided them into 1798 cardiac MRIs for model training. \textbf{BraTS2018}\cite{60} dataset is a 3D brain MRI dataset from 285 patients and each case has four modalities. In this paper, we segmented 3D MRIs under single mode and obtained 6528 2d images for model training and testing. 

\par \textbf{Evaluation Metrics.} In this paper, we use the Fre'chet Inception distance (FID) \cite{61} as the main metric. The FID  is a comprehensive metric, which has been shown to be more consistent with human evaluation in assessing the realism and variation of the generated images. In all our experiments, FID is computed using synthetic images with an equal number of real images. We have fully trained each model and selected the best performance as the experimental results. 

\par \textbf{Baselines.} In this paper, we implemented six unconditional image synthesis models as baselines, including four models proposed or improved in the medical field, namely InfoGAN\cite{71}, DCGAN\cite{86}, WGAN\cite{69}, and DDPM\cite{70}. There are two models in the non-medical field, namely SAGAN\cite{28}, and UNetGAN\cite{30}. Although SAGAN and UNetGAN are aimed at natural image synthesis, they are the main reference networks for this work. Among them, InfoGAN was proposed in 2017 for learning cell-level visual representation; DCGAN was proposed for retinal image synthesis in 2022; WGAN was proposed for brain MRI image synthesis in 2018; DDPM was proposed in 2023 and applied to synthesize a wide range of histopathological images. The original models of these methods were trained on natural datasets, but we chose their improved methods on medical image synthesis as the baselines. 

\par \textbf{Implementation.} 
All methods involved in this experiment  refer to the original paper or the open source code on the network. The generator networks of HAGAN and UNetGAN both add a visual attention mechanism. In order to visually contrast the effect, the generator structure is consistent with SAGAN. At the same time, DCGAN removes the attention mechanism on the basis of SAGAN, and the overall architecture is consistent with the original paper. Finally, all horizontal comparison networks except HAGAN use basic image augmented strategies such as flipping and panning in the data preprocessing to enhance the network effect. In low-resolution experiments, we use a uniformly distributed noise vector $z \in [-1, 1]^{100}$ as input to the generator, and the Adam optimizer \cite{62} with learning rates of $1e-3$ for $G$ and $D$. In high-resolution experiments, the noise vector $z \in [-1, 1]^{1000}$, and the learning rates are $1e-3$. The number of warmup epochs $n$ for consistency regularization is set to 100 for all datasets. Due to computing resource constraints, the mini-batch sizes: 32 for low-resolution and 2 for high-resolution.

\begin{table*}[]
\renewcommand\arraystretch{1.8}
\caption{The Params and Flops of Models.}
\setlength{\tabcolsep}{8.3mm}{
\begin{tabular}{ccccccc}
\hline
\rowcolor[HTML]{FFFFFF} 
\cellcolor[HTML]{FFFFFF}{\color[HTML]{000000} }                                  & \multicolumn{3}{c}{\cellcolor[HTML]{FFFFFF}{\color[HTML]{000000} \textbf{Params}}}                          & \multicolumn{3}{c}{\cellcolor[HTML]{FFFFFF}{\color[HTML]{000000} \textbf{Flops}}}                           \\ \cline{2-7} 
\rowcolor[HTML]{FFFFFF} 
\multirow{-2}{*}{\cellcolor[HTML]{FFFFFF}{\color[HTML]{000000} \textbf{Models}}} & {\color[HTML]{000000} \textbf{G}} & {\color[HTML]{000000} \textbf{D}} & {\color[HTML]{000000} \textbf{ALL}} & {\color[HTML]{000000} \textbf{G}} & {\color[HTML]{000000} \textbf{D}} & {\color[HTML]{000000} \textbf{ALL}} \\ \hline
\cellcolor[HTML]{FFFFFF}InfoGAN\cite{71}                                          & 10.1M                              & 0.1M                              & 10.2M                                & 6.26G                             & 6.21G                             & 12.47G                              \\
\cellcolor[HTML]{FFFFFF}DCGAN\cite{86}                                           & 3.6M                              & 2.8M                              & 6.4M                                & 6.27G                             & 6.19G                             & 12.46G                              \\
\cellcolor[HTML]{FFFFFF}WGAN\cite{69}                                          & 13.3M                              & 6.4M                              & 19.7M                                & 0.79G                             & 0.38G                             & 1.17G                              \\
\cellcolor[HTML]{FFFFFF}SAGAN\cite{28}                                                    & 3.6M                              & 2.8M                              & 6.4M                                & 11.96G                            & 6.19G                             & 18.15G                              \\
\cellcolor[HTML]{FFFFFF}UNetGAN\cite{30}                                         & 3.6M                              & 6.2M                              & 9.8M                                & 11.96G                            & 16.75G                            & 28.71G                              \\
\cellcolor[HTML]{FFFFFF}\textbf{HAGAN(Ours)}                                         & \textbf{3.6M}                              & \textbf{2.8M}                              & \textbf{6.4M}                                & \textbf{11.96G}                            & \textbf{6.19G}                             & \textbf{18.15G}                              \\ \hline
\end{tabular}}
\end{table*}

\begin{table*}[]
\renewcommand\arraystretch{1.8}
\caption{Results of medical image synthesis on 64 × 64 resolution.}
\setlength{\tabcolsep}{6.7mm}{
\begin{tabular}{ccccccc}
\hline
\textbf{Models} & \textbf{Dataset A}                                                        & \textbf{FID$\downarrow$} & \textbf{Dataset B}                                                     & \textbf{FID$\downarrow$} & \textbf{Dataset C}                                                          & \textbf{FID$\downarrow$} \\ \hline
InfoGAN\cite{71}   & \multirow{8}{*}{\begin{tabular}[c]{@{}c@{}}COVID-CT\\ (396 Imgs)\end{tabular}} & 139.679      & \multirow{8}{*}{\begin{tabular}[c]{@{}c@{}}ACDC\\ (1798 Imgs)\end{tabular}} & 110.403      & \multirow{8}{*}{\begin{tabular}[c]{@{}c@{}}BraTS2018\\ (6528 Imgs)\end{tabular}} & 127.563      \\
DCGAN\cite{86}     &                                                                           & 110.793      &                                                                        & 105.316      &                                                                             & 147.805      \\
WGAN\cite{69}      &                                                                           & 124.723      &                                                                        & 97.562       &                                                                             & 76.854       \\
DDPM\cite{70}      &                                                                           & 95.984       &                                                                        & 78.098       &                                                                             & 36.945       \\
SAGAN\cite{28}     &                                                                           & 104.542      &                                                                        & 75.912       &                                                                             & 80.421       \\
UNetGAN\cite{30}   &                                                                           & 100.058      &                                                                        & 103.682      &                                                                             & 82.433       \\
\textbf{HAGAN(Ours)}  &                                                                           & \textbf{88.852}       &                                                                        & \textbf{60.792}       &                                                                             & \textbf{32.551}       \\ \hline
\end{tabular}}
\end{table*}

\subsection{Experimental Results}
\par \textbf{Params and Flops.} In this paper, we use lightweight components to implement HAGAN. By utilizing exquisite structural design, it not only has fast inference speed, but also has better inference ability and generative diversity in terms of pathological structural integrity and local detail continuity. In this test, we will use parameter magnitude to measure the generalization performance limit of the model, and use Flops to measure the inference speed. We pursue higher generalization performance under the premise of faster inference. We tested HAGAN and six baselines, and the results are shown in Table 1. Due to the special nature of the diffusion model, it was not split into two structures, generator and discriminator, and placed in the table. Its parameter magnitude is 35.7M. In the test results, HAGAN has the smallest parameter magnitude, almost no difference from SAGAN. At the same time, the inference speed of our model ranks third among all networks, only behind DCGAN and WGAN, but the overall difference is not significant, and it is significantly better than large networks such as DDPM and other models. In subsequent experiments, we can explain the experimental phenomenon based on the above test results, and find that HAGAN has remarkable advantages in training cost and model accuracy.

\begin{figure*}[htp]
    \centering
    \includegraphics[width=17.7cm]{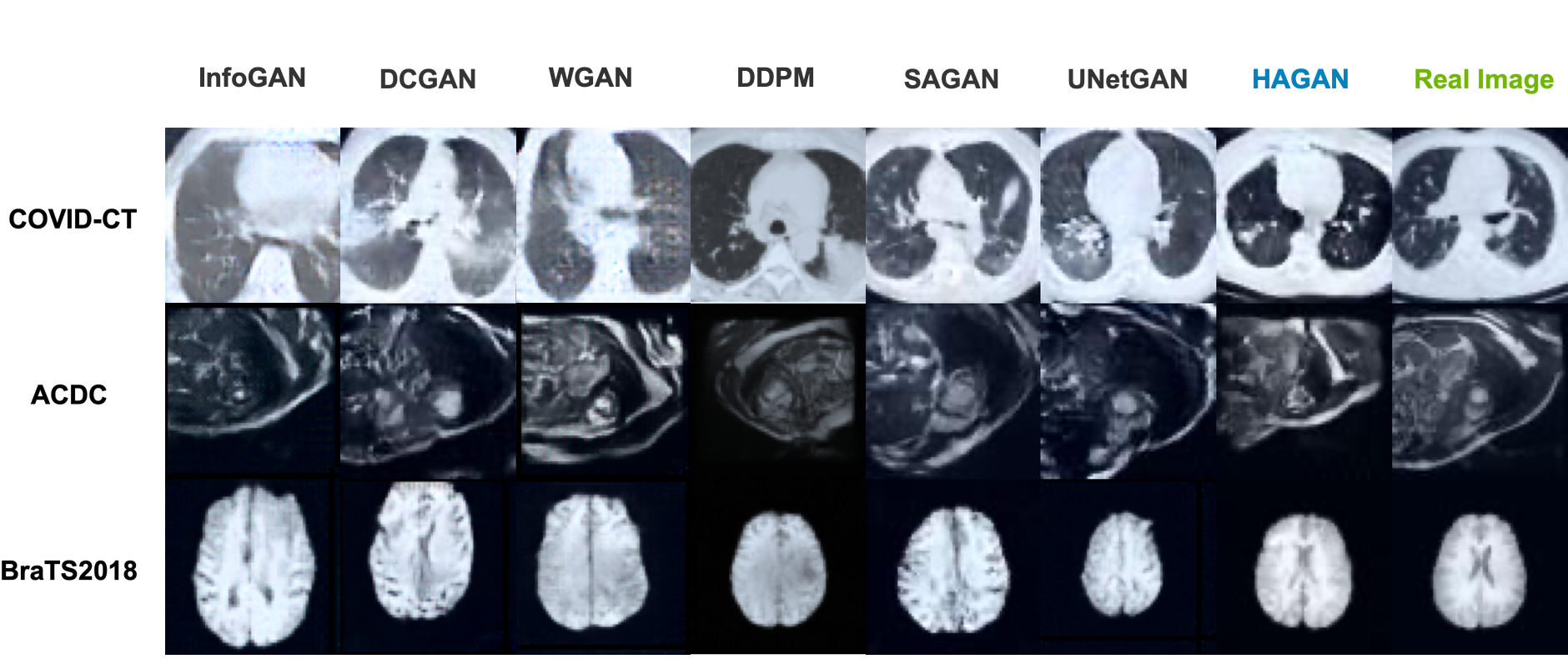}
    \caption{Medical image synthesis samples on 64 × 64 resolution. }
    \label{fig:4}
\end{figure*}

\par \textbf{Low Resolution. } Due to the extensive development of mobile CT technology, the recognition ability of pathological features at low resolution has become a bottleneck that models are eager to break through, but low-resolution image acquisition also requires high costs. To address this problem, we adapted the models to lower resolution images and tested its performance.

\par We preprocessed the data on the three datasets of COVID-CT, ACDC, and BraTS2018 into 64 × 64 resolution. Here we mainly compare the generation capabilities of six baselines and the HAGAN proposed in this paper. The results are shown in Table 2, and Figure 5 is a visualization result. And it should be noted that the goal of this task is to learn the data distribution characteristics of real images, rather than to keep the two completely consistent. Therefore, the visualization result is more of a schematic image selected from numerous real and fake images, providing a more intuitive understanding of changes in data indicators.

\par On the COVID-CT small-scale lung dataset, the FID of HAGAN is 88.852. Among the baselines in the medical field, the DDPM proposed for the histopathology image synthesis in 2023 achieved the best performance with FID of 95.984. However, it is still slightly inferior to HAGAN in this small-scale pneumonia dataset, and the parameter scale is more than six times that of HAGAN. After DDPM, the indicators of models improved on medical image synthesis such as InfoGAN, DCGAN, and WGAN are slightly inferior to those proposed on natural image synthesis such as SAGAN and UNetGAN. We speculate that the main reason is the data generality of medical models is poor. In the new data distribution, with the small-scale dataset, it cannot achieve performance similar to the original task. At the same time, the publication time of these original models is slightly earlier than SAGAN and UNetGAN. Quantitatively speaking, compared with the optimal baseline, HAGAN has an improvement of 7.132 in indicators. And compared to other baselines, there has been a significant improvement of more than 12. Qualitatively speaking, the visualization results show that the images generated by HAGAN have also improved in terms of clarity, texture consistency, and pathological structural integrity.

\par On the ACDC medium-scale heart dataset, the FID of HAGAN is 60.792. Our model has an improvement of 15.12 compared to the optimal SAGAN's FID of 75.912, and an improvement of 17.306 compared to the optimal medical field model DDPM's FID of 78.098. It can be found that SAGAN has better performance on this dataset compared to medical models. We speculate that this is mainly due to the attention mechanism. The generator structure of SAGAN incorporates the visual attention mechanism, which is conducive to perceiving the connection between pixels in cardiac pathological images, better extracting lesion features, and synthesizing higher quality pathological images. The experimental results on the visual attention mechanism are in line with our expectations, which is one of the main reasons why we integrated it into the generative structure of HAGAN. Quantitatively looking at the experimental results, HAGAN has an average index improvement of more than 20 compared to most baselines, while also having a smaller parameter magnitude and faster inference speed. Qualitatively looking at the visualization results, due to the attention mechanism of generator and the pixel-level discriminant feedback introduced by the discriminator, our model has stronger local detail generation ability and better synthesis effect on the details of medical pathological images.

\begin{figure*}[htp]
    \centering
    \includegraphics[width=18.1cm]{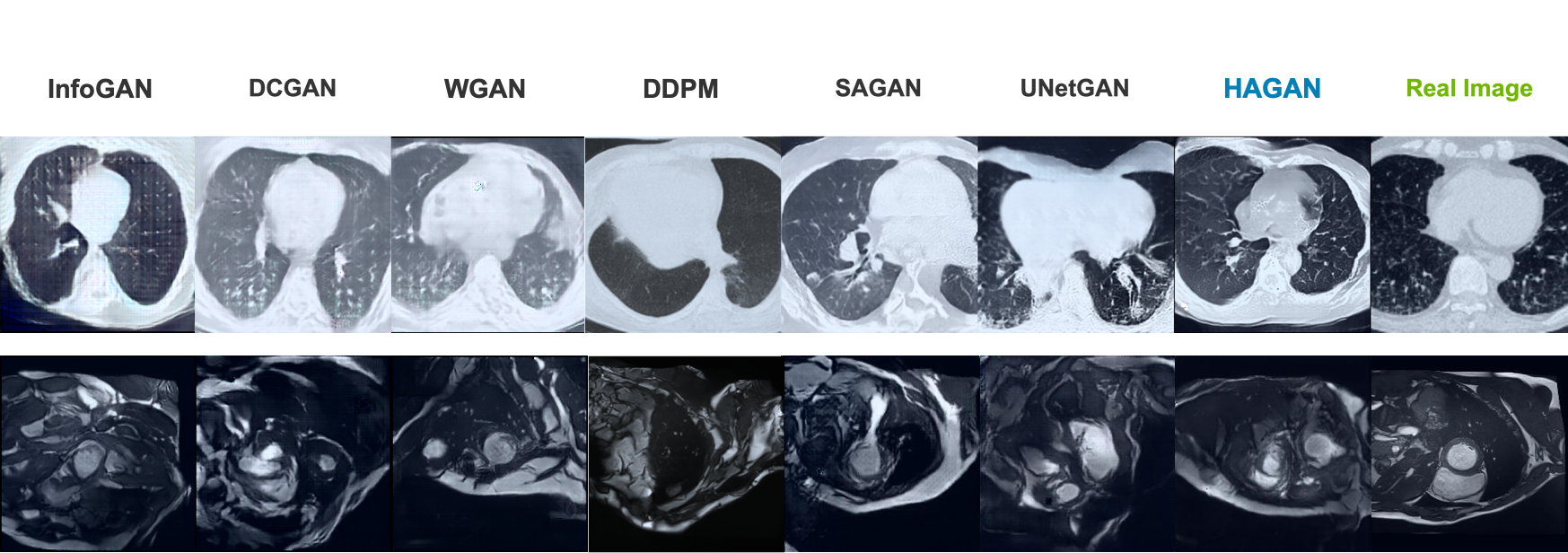}
    \caption{Medical image synthesis samples on 256 × 256 resolution. The top is in COVID-CT dataset and the bottom is in ACDC dataset. }
    \label{fig:5}
\end{figure*}

\par On the BraTS2018 large-scale brain MRI dataset, the FID of HAGAN is 32.551, which is an increase of 4.394 compared to the best-performing medical field model DDPM's FID of 36.945. Compared with the main reference baseline network SAGAN's FID of 80.421, it is a significant increase of 47.87. Compared with UNetGAN's FID of 82.433, it increases 49.882. The reason why the experimental effect of DDPM is great on this dataset, we speculate that it is mainly due to the large size of the dataset, which can better play the generalization potential of large parameter models, thus achieving better generation effect. Quantitatively looking at the experimental results, HAGAN has achieved the best performance in all datasets, which can prove wide applicability and better generalization performance in medical data. Qualitatively, HAGAN has the most complete pathological structure and almost identical detail texture with real images.

\par In low-resolution experiments, HAGAN achieved the best performance on three datasets of different scales and locations. The integrity of its structure and the fineness of local details can also be clearly observed in the visualization results. Compared with the known baselines, its excellent performance at low resolution is more conducive to its adaptation to mobile CT technology, helping smart medical to complete the portable and lightweight transformation. In the future, we will also complete the experiment at high resolution to test the generation effect of large-scale medical pathology images.

\begin{table}[]
\renewcommand\arraystretch{1.8}
\caption{Results of medical image synthesis on 256 × 256 resolution.}
\setlength{\tabcolsep}{2.3mm}{
\begin{tabular}{ccccc}
\hline
\textbf{Models} & \textbf{Dataset A}                                                        & \textbf{FID$\downarrow$}    & \textbf{Dataset B}                                                     & \textbf{FID$\downarrow$}     \\ \hline
InfoGAN\cite{71}           & \multirow{5}{*}{\begin{tabular}[c]{@{}c@{}}COVID-CT\\ (396 Imgs)\end{tabular}} & 197.215         & \multirow{5}{*}{\begin{tabular}[c]{@{}c@{}}ACDC\\ (1798 Imgs)\end{tabular}} & 135.037          \\
DCGAN\cite{86}           &                                                                           & 195.417         &                                                                        & 133.973          \\
WGAN\cite{69}           &                                                                           & 186.094         &                                                                        & 126.731          \\
DDPM\cite{70}           &                                                                           & 182.405         &                                                                        & 112.165          \\
SAGAN\cite{28}           &                                                                           & 139.371         &                                                                        & 112.788          \\
UNetGAN\cite{30}         &                                                                           & 145.617          &                                                                        & 124.861          \\
\textbf{HAGAN(Ours)}  &                                                                           & \textbf{125.871} &                                                                        & \textbf{111.805} \\ \hline
\end{tabular}}
\end{table}

\par \textbf{High Resolution.} In order to prove that the HAGAN method proposed in this paper is suitable for the mainstream medical image resolution and is equally effective in high-resolution imaging, we conducted high-resolution experiments on the COVID-CT small-scale lung dataset and the ACDC large-scale heart dataset. All preprocessing of the data except for resolution is consistent with the low-resolution experiment to ensure the objectivity and effectiveness of the experiment. This experiment still uses four methods proposed or improved in the medical field, InfoGAN, DCGAN, WGAN, DDPM, and two natural image synthesis networks, SAGAN, UNetGAN, as the baselines, and completes high-resolution comparative experiments with HAGAN. The experimental results are still explained from the perspectives of quantitative indicators and qualitative results, and theoretical explanations are given for the changes in indicators and visualization effects. The results are shown in Table 3, and Figure 6 is a visualization result.

\par On the COVID-CT lung small-scale dataset, the FID of HAGAN is 125.871, an increase of 13.5 compared to the best-performing SAGAN's FID of 139.371. Compared with the best-performing medical field model DDPM's FID of 182.405, a significant improvement of 56.534. Compared with the low-resolution experiments on this dataset, the overall performance of the medical field models has declined. It is speculated that the InfoGAN, DCGAN, and WGAN original models were proposed earlier, when the mainstream medical image resolution was 64 × 64, so the adaptability to high-resolution images was poor. Among them, the performance degradation of DCGAN is the most obvious, with FID of 195.417. In addition to the above factors, this is also related to the small parameter magnitude of DCGAN. Expanding the resolution leads to an increase in image features that need to be learned, and the learning difficulty of the model will also increase accordingly. DDPM adopts the principle and structure of Diffusion, and the parameter magnitude is large. Quantitatively looking at the experimental results, whether it is the baseline proposed in the medical field or the natural image synthesis, HAGAN has achieved better performance indicators. Compared with the vast majority of baselines, FID has improved by more than 20. At the same time, even for high-resolution synthesis, the existing small parameter level of HAGAN can complete it, and it has faster reasoning speed. Qualitatively looking at the visualization results, the generation advantage of HAGAN is more obvious, and the integrity of the pathological structure and local detail consistency of its synthesized image can be clearly observed, which is almost no different from the real images.

\begin{table*}[]
\renewcommand\arraystretch{2.0}
\caption{Ablation studies of HAGAN on COVID-CT dataset.}
\setlength{\tabcolsep}{3.9mm}{
\begin{tabular}{clccccc}
\hline
\textbf{Dataset}         & \multicolumn{1}{c}{\textbf{Modules}}             & \textbf{Adv Img Loss} & \textbf{Adv Pix Loss} & \textbf{Conv Img Loss} & \multicolumn{1}{l}{\textbf{Conv Pix Loss}} & \textbf{FID$\downarrow$}                        \\ \hline
\multirow{5}{*}{COVID-CT} & Base Model                                       & $\sqrt{}$             & $\textbf{}$             & $\textbf{}$              & $\textbf{}$                                  & 240.231                             \\
                          & + AttnMix Generator                              & $\sqrt{}$             & $\textbf{}$             & $\sqrt{}$              & $\textbf{}$                                  & 123.270                             \\
                          & + Reverse Skip Connection                        & $\sqrt{}$             & $\textbf{}$             & $\sqrt{}$              & $\textbf{}$                                  & 115.912                             \\
                          & \multicolumn{1}{c}{+ Hierarchical Discriminator} & $\sqrt{}$             & $\sqrt{}$             & $\sqrt{}$              & $\sqrt{}$                                  & \multicolumn{1}{l}{99.717}          \\
                          & + Training Strategy(Ours)                        & $\sqrt{}$             & $\sqrt{}$             & $\sqrt{}$             & $\sqrt{}$                                  & \multicolumn{1}{l}{\textbf{88.852}} \\ \hline
\end{tabular}}
\end{table*}

\par On the ACDC heart large-scale dataset, the FID of HAGAN is 111.805, which is an improvement of 0.36 compared to the best-performing DDPM's FID of 112.165. Although HAGAN has only a slight improvement compared to DDPM, this improvement is huge in the parameter level of the reference model, because the parameter magnitude of DDPM is more than six times that of HAGAN. In addition, among the models proposed on natural image synthesis, SAGAN has excellent performance with FID of 112.788, which is consistent with its low-resolution experimental results on this dataset, further demonstrating the speculation at that time that the visual attention mechanism is conducive to perceiving the connection between pixels in cardiac pathological images, better extracting lesion features, and synthesizing higher-quality lesion images. The experimental feasibility and theoretical effectiveness of the HAGAN structure have been further demonstrated. Quantitatively looking at the experimental results, HAGAN has improved the FID index by more than 10 compared to most baselines on large-scale medical datasets. Qualitative visualization results. The composite images from HAGAN are closer to real images and have better effects on local lesion detail processing. 
\par In summary, we first tested the parameter magnitude and inference speed of HAGAN and six baselines to reflect the model size and inference cost. Based on the experimental results, we conducted 64 × 64 and 256 × 256 resolution experiments around three different sizes of lung, heart, and brain medical public datasets: small, medium, and large. We found that whether it is quantitative indicators or qualitative visualization results, HAGAN has achieved the best performance and the best pathological image generation quality under the premise of combining the advantages of model size and inference speed, fully proving the effectiveness of HAGAN.

\subsection{Ablation Study}
\par We conduct ablation experiments for HAGAN to verify the effectiveness of each module, and the generalization performance improvement brought by each component on the 64 × 64 resolution COVID-CT dataset, and finally calculate FID. Please note that the addition of each individual component here is based on the previously added components, and the components are built on each other. As shown in Table 4, we do not use any data augmented strategy for the basic model, and the structure is consistent with the DCGAN framework without adding any additional network modules. 
\par On this basis, we first used the AttnMix consistency differentiable regularization technology to change the generator structure of the model to AttnMix Generator, which fundamentally improved the generation effect of the model. Quantitatively, the FID of the model increased from 240.331 to 123.270, an increase of 117.061. Qualitatively, the generation effect has also been revolutionized, and the generated pathological images have a complete organizational structure. Whether it is the performance improvement on small-scale datasets or the more complete pathological structure, these effects are consistent with the development expectations of AttnMix technology. Afterwards, we improved the intermediate structure between the generator and discriminator. The feature distributions of the generator and discriminator of the original model do not affect each other, and two independent models were trained separately. In HAGAN, we added Reverse Skip Connection between the two, adding the feature distribution extracted from the real image discriminant path to the feature distribution of the generated path to accelerate the convergence of the model and improve the authenticity of the generation. The addition of this component increased FID from 123.270 to 115.912, and significantly improved the training speed and stability. Subsequently, we improved the structure of the discriminator, adding pixel-level true-fake discrimination on the basis of the original image true-fake discrimination, further constraining the discriminator to learn the key characteristics of local details. This improvement raised FID to 99.717, enhancing the generation ability of local texture and details on the basis of the existing complete lesion structure. Finally, in order to better utilize the full capabilities of AttnMix differentiable regularization technology, we designed a training strategy that integrates global prior information, greatly improving the stability of training and further raising FID to 88.852.

\begin{figure}[htp]
    \centering
    \includegraphics[width=9.0cm]{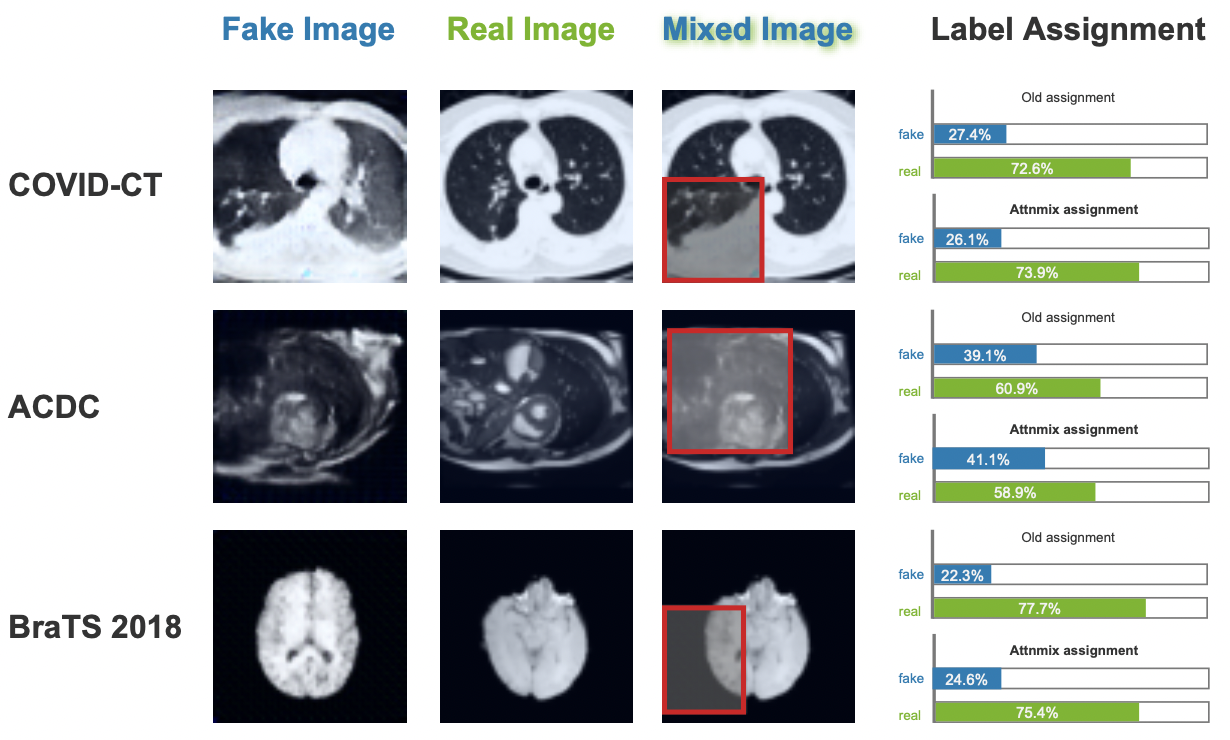}
    \caption{Visualization of AttnMix consistent differentiable regularization. }
    \label{fig:6}
\end{figure}

\subsection{AttnMix Visualization}
\par We visualized the true and fake images fusion process and label allocation weights of AttnMix consistency differentiable regularization technology on three medical public datasets, as shown in Figure 7. These samples are all from the real training process. The first and second columns show the fake and true images used in the training process, and the two images are mixed in the third column. The fourth column shows the label allocation strategy based on the mixing region and the label allocation ratio based on saliency. On the COVID-CT dataset, we fused the lower left part of the composite image to the corresponding position of the real image. If the label of the mixed image is calculated according to the area ratio, the authenticity will be 0.726. However, in fact, some pixels in the replaced lower left image cannot be used to distinguish whether it is a real image, and it is unreasonable for these pixels to participate in the label proportion calculation. At the same time, even if pixels really play a role in true and fake discrimination, the intensity of this feature is worth further judgment and affects the label proportion allocation. In AttnMix, we use the attention map generated during the generation process to determine the intensity of replacement features and further affect the proportional allocation of labels, so that the model can more accurately focus on those non-class preservation perturbations, thereby enhancing the ability of extracting key features of tissue cells or lesion sites. After adjustment by AttnMix, the authenticity of the mixed image is 0.739, and this value will be used as a label to construct the discriminant task of the mixed image. Similarly, after mixing and saliency adjustment on ACDC, the authenticity of the mixed image is corrected from 0.609 to 0.589. The authenticity of the mixed image on the BraTS2018 dataset is corrected from 0.777 to 0.754. Please note that the change here is small because the center of gravity of medical image generation is the integrity of the global pathological structure. The generation task of medical images does not have obvious generative subjects that need to be identified, so the significant weight is relatively scattered. However, subtle weight changes will have a significant impact on the consistency of local details and the integrity of lesion sites, and further impact the authenticity of medical image synthesis.

\section{Conclusion}
In this study, we propse HAGAN, a hybrid augmented generative adversarial network for medical image synthesis. HAGAN incorporates Attention Mixed (AttnMix) Generator, Hierarchical Discriminator, and Reverse Skip Connection bridging the Discriminator ($D$) and the Generator ($G$). The AttnMix Generator employs a consistency differentiable regularization mechanism, encouraging the model to focus more on structural and textural information, thereby boosting its ability to generate complete pathological structures and intricate local details. Simultaneously, the Hierarchical Discriminator provides discriminant feedback at both image and pixel bi-level for the generator, thereby improving the saliency and discriminability of the model. The Reverse Skip Connection guarantees the authenticity of synthetic images by integrating the features extracted from the true discrimination pathway into the generated feature maps. Finally, HAGAN has demonstrated enhanced performance at both the high-resolution and low-resolution experiments for medical image synthesis on three different scale datasets, i.e., COVID-CT, ACDC and BraTS2018.

\bibliographystyle{IEEEtran}

\bibliography{related_work}

\end{document}